\documentclass[conference]{IEEEtran}

\usepackage{amsmath,amssymb,amsfonts}
\usepackage{algorithmic}
\usepackage{amsmath,amssymb,amsfonts}
\usepackage{algorithmic}
\usepackage{subcaption}
\usepackage{graphicx}
\usepackage{textcomp}
\usepackage{xcolor}
\usepackage{float}
\usepackage{longtable}
\usepackage{tabulary}
\usepackage{tabularx}
\usepackage{dblfloatfix}
\usepackage{outlines}
\usepackage{placeins}
\usepackage{booktabs}
\usepackage{color, colortbl}

\usepackage{multirow}
  
\usepackage{tikz}
\usepackage[style=ieee]{biblatex}
\usepackage[flushleft]{threeparttable}

\usepackage[english]{babel}
\usepackage{fancyhdr}
\usepackage{lastpage}

\usepackage{booktabs}

\def\BibTeX{{\rm B\kern-.05em{\sc i\kern-.025em b}\kern-.08em
    T\kern-.1667em\lower.7ex\hbox{E}\kern-.125emX}}

\definecolor{myred}{RGB}{255, 10, 10}

\usepackage{enumitem}

\begin{document}
\bstctlcite{IEEEexample:BSTcontrol}



\title{FlowTransformer: A Transformer Framework for Flow-based Network Intrusion Detection Systems}


\author{
\IEEEauthorblockN{
Liam Daly Manocchio\IEEEauthorrefmark{1},
Siamak Layeghy\IEEEauthorrefmark{2},
Wai Weng Lo\IEEEauthorrefmark{3},
Gayan~K.~Kulatilleke\IEEEauthorrefmark{4},
Mohanad Sarhan\IEEEauthorrefmark{5},\\
Marius Portmann\IEEEauthorrefmark{6}
 \\
}
\IEEEauthorblockA{
\textit{School of Information Technology and Electrical Engineering} \\
\textit{University of Queensland}\\
Brisbane, QLD 4072, Australia\\
\IEEEauthorrefmark{1}liam@riftcs.com, \IEEEauthorrefmark{2}siamak.layeghy@uq.net.au,
\IEEEauthorrefmark{3}w.w.lo@uq.net.au,
\IEEEauthorrefmark{4}g.kulatilleke@uqconnect.edu.au,\\
\IEEEauthorrefmark{5}m.sarhan@uq.net.au,
\IEEEauthorrefmark{5}marius@ieee.org
}
}

\IEEEoverridecommandlockouts
\makeatletter\def\@IEEEpubidpullup{6.5\baselineskip}\makeatother
\IEEEpubid{\parbox{\columnwidth}{
    Network and Distributed System Security (NDSS) Symposium 2024\\
    26 February - 1 March 2024, San Diego, CA, USA\\
    ISBN 1-891562-93-2\\
    https://dx.doi.org/10.14722/ndss.2024.23xxx\\
    www.ndss-symposium.org
}
\hspace{\columnsep}\makebox[\columnwidth]{}}

\maketitle

\begin{abstract}

This paper presents the FlowTransformer framework, a novel approach for implementing transformer-based Network Intrusion Detection Systems (NIDSs). FlowTransformer leverages the strengths of transformer models in identifying the long-term behaviour and characteristics of networks, which are often overlooked by most existing NIDSs. 
By capturing these complex patterns in network traffic, FlowTransformer offers a flexible and efficient tool for researchers and practitioners in the cybersecurity community who are seeking to implement NIDSs using transformer-based models. FlowTransformer allows the direct substitution of various transformer components, including the input encoding, transformer, classification head, and the evaluation of these across any flow-based network dataset.
To demonstrate the effectiveness and efficiency of the FlowTransformer framework, we utilise it to provide an extensive evaluation of various common transformer architectures, such as GPT 2.0 and BERT, on three commonly used public NIDS benchmark datasets. We provide results for accuracy, model size and speed.
A key finding of our evaluation is that the choice of classification head has the most significant impact on the model performance. Surprisingly, Global Average Pooling, which is commonly used in text classification, performs very poorly in the context of NIDS.
In addition, we show that model size can be reduced by over 50\%, and inference and training times improved, with no loss of accuracy, by making specific choices of input encoding and classification head instead of other commonly used alternatives.

\end{abstract}


\section{Introduction}

ChatGPT~\cite{openai2022chatgpt} has seen an explosion in use, in academic, private, and professional domains, for its ability to answer complex prompts and generate high-quality human-like text. 
ChatGPT, like many other large language models, is based on a transformer architecture. Transformer architectures are extremely powerful for natural language processing (NLP) tasks due to their ability to capture long-range dependencies and relationships between different elements of a sequence, without requiring prior domain-specific knowledge or feature engineering~\cite{vaswani2017attention}. 
Although initially designed for NLP, transformer architectures have proven to be versatile and powerful tools for capturing complex patterns and relationships in various types of sequential data, including but not limited to image, graph, and speech ~\cite{vaswani2017attention,parmar2018image,yun2019graph}. This adaptability has made transformer-based models particularly attractive for use in machine learning (ML)-based Network Intrusion Detection Systems (NIDSs), where data is captured as sequences of packets or flows, and where the ability to identify subtle and complex patterns in this traffic is critical for NIDS performance. 

Despite the sequential nature of network communications, current ML-based NIDS research often overlooks sequential data, focusing instead on classifying individual network flow records in isolation. This is partly due to the challenges posed by traditional ML models in handling sequential data.
Transformer architectures, on the other hand, provide an effective means of applying machine learning to a sequence of data. 
In addition, detecting a large proportion of attacks in network traffic requires considering the long-term behavior and characteristics of the network.
The transformers' ability to access long-range information can help detect complex patterns of network traffic that may indicate attacks, even if they are distributed over an extended period of time. 
Therefore, we believe there is a great potential in investigating transformer architectures such as GPT~\cite{openai-gpt2-1.5b} and BERT~\cite{devlin2018bert}, or shallower transformers, in the context of NIDS.

Applying transformers in the networking domain is not as straightforward as in the natural language domain. Although established architectures exist for handling text data with transformers, this is not the case for network traffic. There are several critical decisions that must be made, independently of the transformer model itself, both to ingest network data, as well as produce a classification from the transformer output. To the best of our knowledge, there are no works that provide an extensive and systematic evaluation of transformer models and parameters in the context of NIDS. 
To address this gap, this paper proposes `FlowTransformer', a framework that allows comprehensive evaluation of transformer-based NIDSs, by enabling the efficient interchange of key components, such as input encoding, transformer model and classification head.     
%

We choose to use flow-based network data for this paper for several reasons. 
Firstly, flow records aggregate network data and provide a compact representation of network communication between two endpoints. This reduces the volume of data that needs to be analysed, making it more scalable for NIDS to process large amounts of network traffic. A significant fraction of networks are of a scale where packet captures are impractical or impossible. Furthermore, most networks are already running flow-based traffic collectors, making it a practical data format for NIDSs.

The key contributions of this paper are that we propose the FlowTransformer framework, to address the challenges associated with implementing transformer-based NIDSs in a systematic manner, and provide the implementation for public use\footnote{FlowTransformer source code https://github.com/liamdm/FlowTransformer}.  It provides researchers the flexibility to select the most appropriate hyperparameters and architecture configurations that align with the requirements of  NIDS applications. 
FlowTransformer provides a systematic evaluation methodology to assess the performance of different transformer models in the context of NIDS. It takes into account model components, hyperparameter configurations, and their impact on model size, speed, and accuracy, allowing for a comprehensive analysis of various transformer architectures in NIDS use cases. 
%
%
Our analysis highlights the trade-offs between model size, performance, and inference time in network intrusion detection tasks, and we demonstrate an over 50\% reduction in model size and a reduction in inference time for the same accuracy when using a sensible choice of transformer components versus other common approaches. 
Finally, as a result of this comparative analysis, we recommend an overall transformer architecture that we believe represents the best choice of components as a starting point for future transformer-based NIDS research. 

\section{Background - NetFlow}


Network traffic monitoring is a critical aspect of network security and management. The two primary  approaches for this purpose include packet-based and flow-based monitoring. Packet-based monitoring involves capturing both packet headers and payloads as they traverse the network, while flow-based monitoring collects summary information based on a sequence of packets between two endpoints. 
However, due to its resource-intensive nature, continuous packet-based monitoring is difficult to implement in large-scale networks. Furthermore, packet capture raises privacy concerns, as it may collect sensitive information. In contrast, flow-based monitoring provides a highly compressed summary of network traffic, making it a more scalable alternative. It is widely used in large-scale networks, and numerous tools are available for flow-based traffic exporting and collection.

NetFlow is a widely adopted flow-based network traffic information collection and monitoring protocol developed by Cisco~\cite{claise2004cisco}.
It operates by consolidating a series of packets within a communication sequence, either unidirectional or bidirectional, that share similar characteristics, such as the same source and destination IP, source, and destination port, and transport protocol. Bidirectional NetFlow is equipped to capture the number of packets and bytes in both directions, along with other features.


Although in this work we use NetFlow version of benchmark NIDS datasets, there are other flow standards, such as IPFix \cite{rfc7011ipfix} or SFlow \cite{rfc3176sflow}, and these flow formats are also natively supported by the FlowTransformer implementation.

\section{Related Works}

This work considers related works that also focus on transformer or RNN based approaches to NIDS. This is because unlike traditional ML-based NIDS that act on single flow, these approaches handle sequences of flows.

Wu et al. \cite{wu2022rtids} present a Transformer-based Intrusion Detection System called RTIDS, which incorporates the positional embedding technique to associate sequential information between features. To train and evaluate the model, a variant stacked Transformer encoder-decoder neural network has been utilised. The approach has been evaluated using the CICDDoS2019 dataset, and its performance has been compared to baseline machine learning models, including support vector machines (SVM), as well as deep learning algorithms such as recurrent neural network (RNN), fuzzy neural network (FNN), and long short-term memory (LSTM). 

Yangmin et al. \cite{li2022extreme} present a novel approach, the Extreme Semi-Supervised Framework based on Transformer (ESeT), for network intrusion detection. This framework utilises a small amount of labeled data to achieve superior detection performance. ESeT includes a multi-level feature extraction module and a semi-supervised learning module, which incorporates a dual-encoding transformer, credibility selector, and feature augmentor. The efficacy of the proposed framework is evaluated on two large, real-world NIDS datasets, and the results demonstrate improved performance compared to existing state-of-the-art methods with only a limited quantity of labeled data.

Wei et al. \cite{wang2023robust} propose a self-supervised-based network intrusion detection system based on transformer. They use a transformer-based architecture and a masked context reconstruction module to detect intrusions. The authors evaluate the algorithm on three different datasets: KDD, UNSW-NB15, CICIDS-17 and also investigate the impact of different parameters of the proposed model, such as the mask ratio and context size, on the algorithm's performance. They show that the algorithm's performance is sensitive to the mask ratio but relatively insensitive to the context size.

Nam et al. \cite{nam2021intrusion} present a novel method for detecting spoofing attacks in Controller Area Networks (CAN) by training a transformer-based language model (GPT) on normal CAN ID sequences. The proposed method outperforms other state-of-the-art methods in terms of accuracy and efficiency. However, the proposed method assumes that the attacker does not have access to the normal CAN ID sequences, which may not be a realistic assumption in practice. 

Loc et al. \cite{nguyen2022flow} propose a method to improve the domain adaptation capability of Network Intrusion Detection Systems (NIDS) by employing the Bidirectional Encoder Representations from Transformers (BERT). The proposed method uses sequences of flows to overcome the limitation of modeling the distribution of features within a flow. The BERT framework is used to utilise the context information from a sequence of flows, allowing the classifier to further model the distribution of a flow in relation to other flows. The BERT model is pre-trained with only the Masked Language Modeling (MLM) task and is fine-tuned with a linear layer with softmax output for detecting intrusion.

Han et al. \cite{HAN2023103171} present a new intrusion detection method for encrypted traffic called GTID. It combines n-gram frequency and time-aware transformer methods to address the limitations of deep learning-based and n-gram-based methods. The model processes packet header and payload features separately, uses n-gram frequency to handle variable-length sessions and packets, and incorporates a time-aware transformer to consider the time intervals between packets. However, the performance of GTID varies with the proportion of encrypted traffic on the dataset, and the method is less effective in detecting encrypted traffic with smaller payload sizes. Also, the use of n-grams results in higher computational complexity.


Mohammad et al. \cite{Mohammad} propose a NIDS (Network Intrusion Detection System) that combines Convolutional Neural Network (CNN) and Long Short-Term Memory (LSTM) architectures using deep learning techniques. The CNN layers are utilised to efficiently extract significant features from network data because of their weight-sharing property, resulting in faster processing speed. The LSTM, on the other hand, maintains the long-term temporal relationship between the extracted data features. The hyper-parameters of the system were optimized through a trial and error process. The hybrid model achieved an overall classification accuracy of 97.1\% for binary classification and 98.43\% for the multi-class case when tested on the UNSW-NB15 dataset.


In contrast to existing works, our proposed framework presents a comprehensive evaluation of transformers in the NIDS domain, systematically exploring the impact of various model components and configurations on performance, model size, and inference time.

\section{Transformers for NIDS}

%
%
As discussed, there has been limited investigation into NIDS that can effectively handle sequences of traffic flows. Transformers provide an attractive option for researchers, given their inherent ability to process sequences and capture complex relationships between items within a sequence. In this context, we will begin by discussing the main components of a transformer-based system when applied to the NIDS task.

\begin{figure}[!t]
    \centering
    \includegraphics[width=1\linewidth]{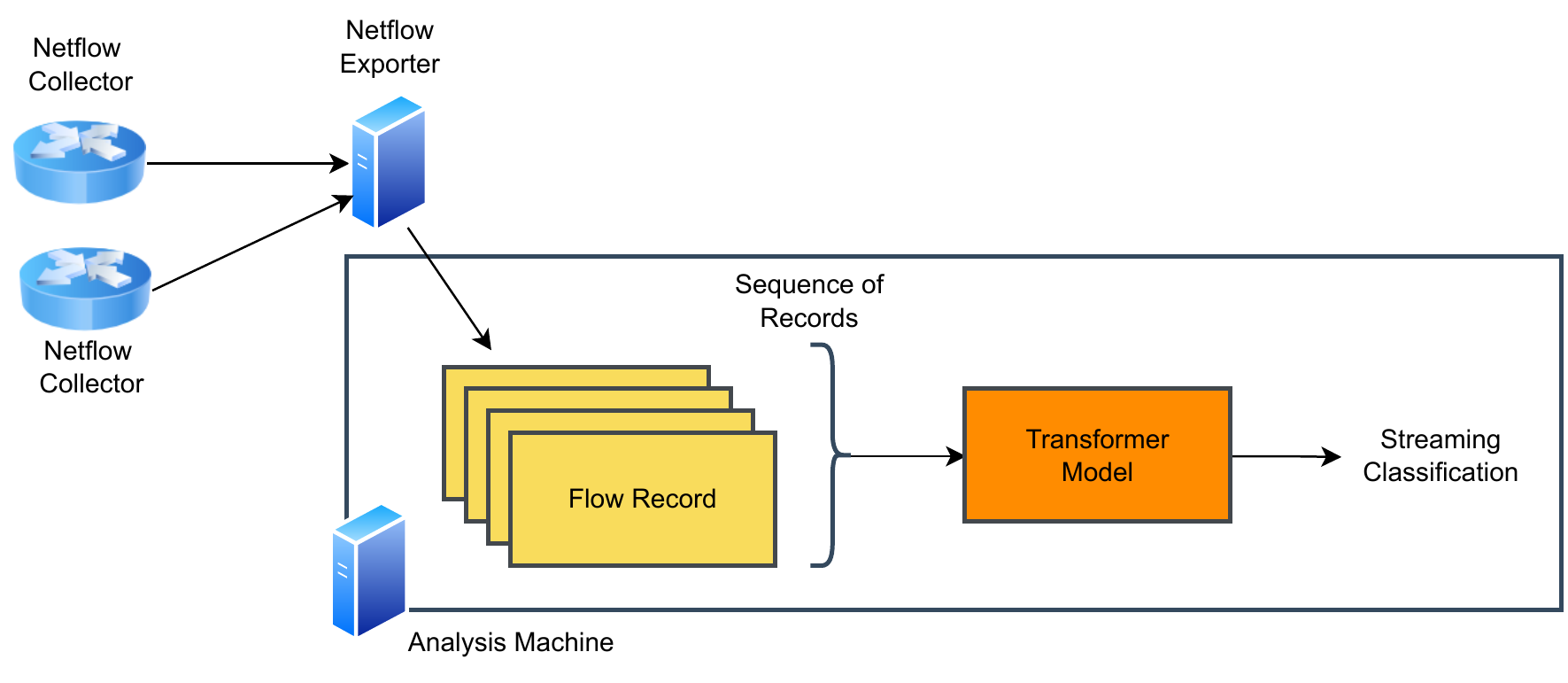}
    \caption{A simple framework for online NetFlow processing using a pre-trained transformer model}
    \label{fig:NetworkBasedFramework}
\end{figure}

Figure~\ref{fig:NetworkBasedFramework} presents a potential framework that employs a transformer-based approach for NetFlow processing. This approach involves using a transformer to analyse sequences of network data. Specifically, the flow records aggregated by a NetFlow exporter are fed directly into a transformer model, which provides classifications for each of these records.

Using a transformer-based approach offers several benefits for this framework. Firstly, it enables the analysis of network data in a highly efficient and scalable manner, given that transformers support parallel execution, unlike many other methods for processing sequential data. Additionally, the transformer model can effectively capture the complex relationships between different network flows, making it ideal for use in NetFlow processing.

However, choosing the right transformer model is crucial. There are several transformer models available, each with unique strengths and weaknesses It is important to consider these differences when selecting a suitable transformer model in the context of network intrusion detection.

\begin{figure}[!b]
    \centering
    \includegraphics[width=1\linewidth]{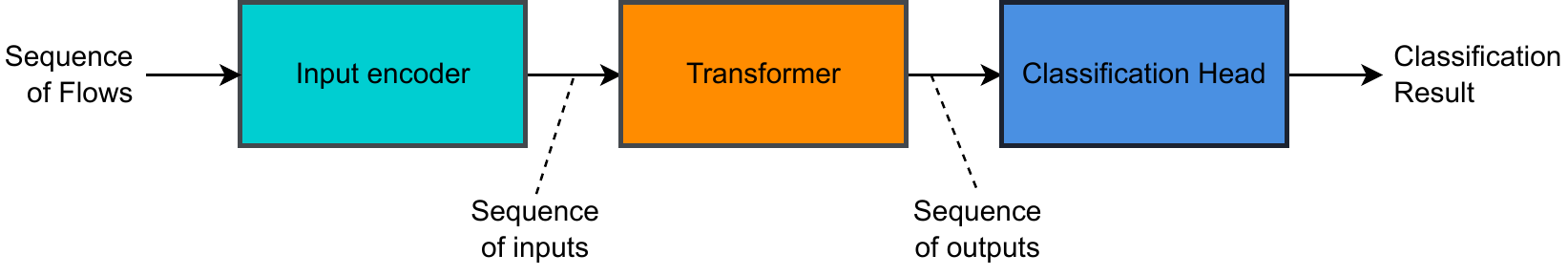}
    \caption{
    Architecture of a transformer-based NIDS consisting of the input encoding, transformer and classification head
    }
    \label{fig:Architecture_Overall_Transformer}
\end{figure}

Transformers are a diverse class of machine learning models that can perform various tasks. For classification or anomaly detection, which are fundamental tasks in NIDS, the basic transformer architecture comprises three components, as illustrated in Figure~\ref{fig:Architecture_Overall_Transformer}. These components include an input encoding module that transforms the input data into a format suitable for use by a transformer, a transformer composed of a number of transformer blocks, and an output head that converts the sequential output of the transformer into a single classification result. 
Each of these steps involves design choices that can significantly impact the model's overall performance. The best choices of these for flow-based data, in contrast to other domains, remains largely un-evaluated.

In this section, we provide a brief explanation of these components within the context of NIDS. Later, we will discuss the options for each of these stages, as implemented in our framework.

\subsection{Input Encoding}
Input encoding for classification is a task similar to tokenisation in natural language processing. In NLP, input encoding involves taking written text as input and transforming it into a sequence of fixed-length vectors that can be processed by a transformer model. For example, the GPT series of models used a variant of byte pair encoding to do this \cite{Takeda1999}. In the context of NIDS, the objective is to transform network flows into a format that can be ingested by a transformer model. This is distinct from pre-processing, as it is a part of the model itself, and the encoding is learned during the model training process.
Although raw data can be provided to a transformer model directly after pre-processing, most transformer implementations have their own input encoding step within the model \cite{vaswani2017attention}. There are various approaches to input encoding discussed in the literature \cite{wolf2020huggingfaces}, and these will be further explored in Section~\ref{FlowTransformerFramework}.

\subsection{Transformer Blocks}

Transformers consist of a sequence of blocks, each of which performs a single `transformation' on the input sequences. These can be encoder or decoder blocks.
The encoder block in a transformer aims to ingest an input and transform this into a fixed-length feature representation. This representation captures the semantic meaning of the particular input, considering its relation to the other inputs in the sequence. In the NIDS domain, the encoder would transform each flow into a fixed-length feature vector.
Decoder blocks are the reverse of an encoder block, and they are more commonly utilised for generational tasks. Decoders take in a sequence of feature representations, and then produce an output sequence. In the NIDS domain, this would take a flow's feature vector, and then produce a raw flow record.

Traditional transformers initially used a stack of encoders, followed by a stack of decoders \cite{vaswani2017attention}.
However, when used for non-sequential tasks, such as classification in NIDS, where a single classification output is generated based on a sequence of flow inputs, the decoder layers can be removed and replaced with a classification head, creating a model built entirely from encoder blocks. This is the approach taken by models such as BERT. BERT was proposed in \cite{devlin2018bert}, as a bidirectional model that could consider tokens in both directions, and solve a number of tasks.
Generative transformer models such as the GPT use the opposite approach, and build a model using exclusively decoder blocks.

\subsection{Classification Head}

Transformers are sequence-to-sequence models, which means the input and output are both sequences. For an NIDS we instead want a classification as output.
The classification head is able to take the sequence of output tokens from the transformer component, and convert these into a prediction over one or more classes.
The primary challenge with classification heads is the dimensionality problem. Transformers can handle effectively arbitrarily long sequences, but if we take the output and attempt to pass this directly to a dense neural network, this could result in an exponentially increasing number of parameters. 
Instead, classification heads often aim to choose particular elements or summarise the information returned from the transformer, so that they do not incur the same dimensionality increase as the sequence length increases.
There has been recent work proposing new classification heads such as \cite{Ridnik_2023_WACV}, which leverage attention mechanisms to process the output of the transformer, however, many traditional approaches used a simple Global Average Pooling approach on the output of the transformer \cite{ko2022group}. Despite the fact that several other basic approaches to classification heads exist, these remain largely unevaluated in the NIDS domain.

\section{FlowTransformer Framework}
\label{FlowTransformerFramework}

FlowTransformer is a framework for implementing, testing and evaluating transformer-based NIDSs. The framework provides researchers with several common implementations for each of the transformer components, forming a pipeline for rapid transformer development that can be directly applied to flow-based networking data.
One of the key advantages of FlowTransformer is its ability to eliminate the difficulties of choosing input encodings and classification heads, which can significantly speed up the research effort. This is particularly important in the NIDS domain, where the ability to practically test certain architectures against a wide range of datasets is paramount.

In the following sections, we will delve into the various components of FlowTransformer, including pre-processing techniques, input encodings, transformer blocks, and classification heads. We will also discuss how FlowTransformer can be extended to implement new components, enabling researchers to develop and test innovative NIDS architectures with ease.

\subsection{Input Encoding Options}

\begin{figure}[!b]
    \centering
    \includegraphics[width=1\linewidth]{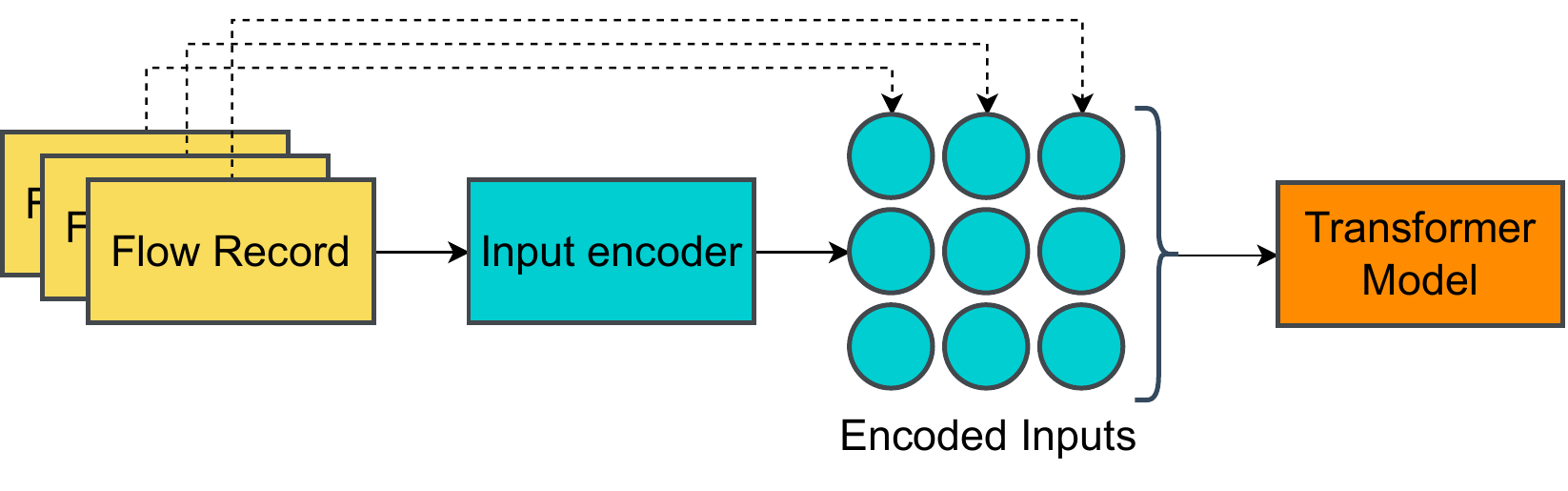}
    \caption{In the NIDS domain, pre-processed flow records are transformed by the input encoder into a fixed length vector, before being passed to the transformer.} 
    \label{fig:Architecture_Encoding_Overall}
\end{figure}

In the context of flow-based NIDS, input encoding is a critical process that allows for the transformation of flow data into a fixed-length feature representation. As shown in Figure~\ref{fig:Architecture_Encoding_Overall} the input encoder is responsible for converting each flow into a feature vector that can be effectively processed by the transformer.
%
Flow data is a type of tabular data, where the fields for each flow are known and fixed. This data consists of two types of fields: numerical fields, such as the number of packets sent, where the distance between numbers has contextual meaning, and categorical fields, such as port number, which have a discrete set of values referred to as levels.
%
With categorical fields, there is no guarantee that the value of a level bears any contextual relevance to other levels. For example, port 25 and 22 are very close but represent distinct protocols, while ports 443 and 8080 are often used for the same service despite their numerical distance. The input encoder must be designed to handle these different types of fields appropriately, ensuring that the resulting feature vectors capture the relevant information for the transformer to use.

%
%
When it comes to transformer-based NIDSs, there are two primary approaches for handling tabular data. One common method, employed by previous works such as "TabTransformer"~\cite{huang2020tabtransformer}, is to encode categorical fields individually by transforming them into a contextually meaningful continuous vector. These vectors are then concatenated with the numerical fields, resulting in a feature vector that can be used by other machine learning models.
While this approach has been successful, it only encodes the categorical features, leaving the numerical fields untreated. An alternative method is to embed all fields simultaneously, allowing for the processing of both numerical and categorical fields together. 
To achieve this, categorical fields can be one-hot encoded and then concatenated with the numerical values and passed to an encoding layer. However, certain fields may be difficult to encode due to the high dimensionality of the resulting feature space. Previous works in the NIDS domain have demonstrated that state-of-the-art performance can still be achieved using frequency-limited one-hot encoding to overcome this increase in dimensionality.

\subsubsection{Categorical Only}

In terms of approaches to encoding categorical fields as part of the input encoding step, there are three main approaches:

\begin{itemize}
    \item \textbf{Lookup-Based Embedding Layer~\cite{mikolov2013distributed}:} The lookup-based embedding layer maps a categorical input into a continuous feature vector. This is implemented as a lookup table, mapping between the unique values of the input, and the set of feature vectors. During training time, the values of these vectors are learned.
    \item \textbf{Fully Connected Embedding Layer:} This is also referred to as a \textit{Dense Embedding Layer}, and it maps a categorical input to a continuous feature vector, using a non-linear transformation. In essence, this is a single layer neural network with an activation function. This transformation is learned during training time, but unlike the lookup-based embedding layer, embeddings are resolved by applying mathematical operations on the input rather than looking up a pre-learned vector.
    \item \textbf{Projection Layer:} The projection layer or linear projection layer, is similar to the dense embedding layer, however, instead of applying a non-linear transformation, the projection layer applies a linear transformation. This can be seen as a very basic form of dimensionality reduction and is widely used for this purpose \cite{sorzano2014survey}.
\end{itemize}

Previous works have proposed the processing of categorical fields with a separate transformer model before concatenating, as a means of embedding categorical fields in tabular data \cite{huang2020tabtransformer}. However, in addition to the large increase in model complexity from such an approach, because this was a non-sequential implementation, built to process a single row of tabular data and make a classification, we do not implement it as part of FlowTransformer.

\subsubsection{Categorical + Numerical - Record Level}

Instead of only embedding the categorical fields, which also requires a transformation to be applied to each categorical field individually, we can instead embed the entire flow record in one pass. This can also be seen as a form of dimensionality reduction, and given that numerical fields in flow records such as number of bytes and number of packets sent often has significant portions of shared information, it is sensible to expect there to be a high level of redundancy in the raw representation of flows that can be removed through this encoding.

The two dense approaches used to embed categorical fields can also be used to embed a flow record. However, lookup-based embedding does not function with (continuous) numerical fields, and thus cannot be applied to the record as a whole which includes both categorical and numerical fields.

\begin{itemize}
    \item \textbf{Record Level Embedding - Dense:} To apply dense embedding, the categorical fields are one-hot encoded and concatenated with the numerical fields before being passed to the embedding layer which then maps the record to a continuous feature vector.
    \item \textbf{Record Level Projection:} The projection layer is implemented in the same manner as the dense record level embedding, omitting the non-linear activation and bias.
\end{itemize}

\subsubsection{Other Approaches}

\begin{itemize}
    \item \textbf{No encoding:} Since the flow records are already pre-processed, it is possible to pass them directly to the transformer if the categorical fields are one-hot encoded. This can significantly increase the dimensionality, but for lightweight flow formats with a small number of features, this is a potentially feasible approach.
    \item \textbf{Handling as text:} This paper does not explore handling flow records as text, for several reasons. Notably, flow data has a structured format, with distinct features. If we were to handle this data as text, this structure is lost, making it harder for the model to capture relationships between features and their corresponding values. Furthermore, transformers are designed to handle sequence data and capture dependencies within a sequence. In fixed-format data, relationships between different fields or elements may not be sequential. This can make it difficult for the transformer to accurately capture these relationships, leading to suboptimal performance.
\end{itemize}

\subsection{Transformer Models}

This study compares four transformer architectures, namely shallow encoder-based, shallow decoder-based, deep encoder-based, and deep decoder-based transformers. The shallow models are based on the basic multi-head self-attention transformer architecture and comprise between 2 and 6 encoder or decoder blocks. Two specific deep transformer models, GPT 2.0 and unmasked BERT, are also considered. The difference between shallow and deep models lies primarily in their depth, number of attention heads, and internal size, while their core transformer block structure is the same. Although both GPT and BERT use scaled-dot-product attention, BERT's attention mechanism is bidirectional, considering tokens in both directions, unlike GPT, which only considers previous tokens.

\subsubsection{GPT 2.0 - Deep Decoder Transformer~\cite{openai-gpt2-1.5b}}

GPT 2.0 is a generative model that is trained on next-word prediction. Unlike traditional input-output models, it treats the input prompt as part of a sequence and generates an output by using each generated word as part of the context to predict the subsequent word. This approach allows GPT to use exclusively transformer decoder blocks, rather than the traditional encoder-decoder structure. The internal block structure is repeated, with smaller GPT 2.0 models having 12 blocks. The input sequence is passed through these decoder blocks one by one, with the output of each block being fed as input to the next block. By using a stack of decoder blocks, GPT can model the distribution of natural language more effectively than a traditional transformer model, because the decoder blocks learn to generate the next word based on a combination of the input sequence and the previously generated words, instead of using only the input sequence. GPT is an autoregressive model that considers tokens only to the left of the token it is generating, moving through a sequence one token at a time.

\subsubsection{BERT - Deep Encoder Transformer~\cite{devlin2018bert}}

BERT, in contrast to GPT, is an encoder-only transformer model, consisting of repeated blocks of transformer encoders. This is because BERT was primarily designed for natural language understanding tasks, rather than text generation.
In BERT, the input sequence is processed by a stack of transformer encoder blocks to generate a fixed-length representation of the sequence. In the case of BERT, the model was trained to perform masked language modelling, as well as predicting if two sentences are consecutive or not. 
This fixed-length representation, however, can be used for a variety of downstream natural language understanding tasks, after fine-tuning the base model.
BERT is not an autoregressive model, and is capable of using the entire surrounding context to generate representations. This means the generated representations depend on the tokens to both the 'left' and 'right'.

\subsection{Classification Head Options}

\begin{figure}[!t]
    \centering
    \includegraphics[width=1\linewidth]{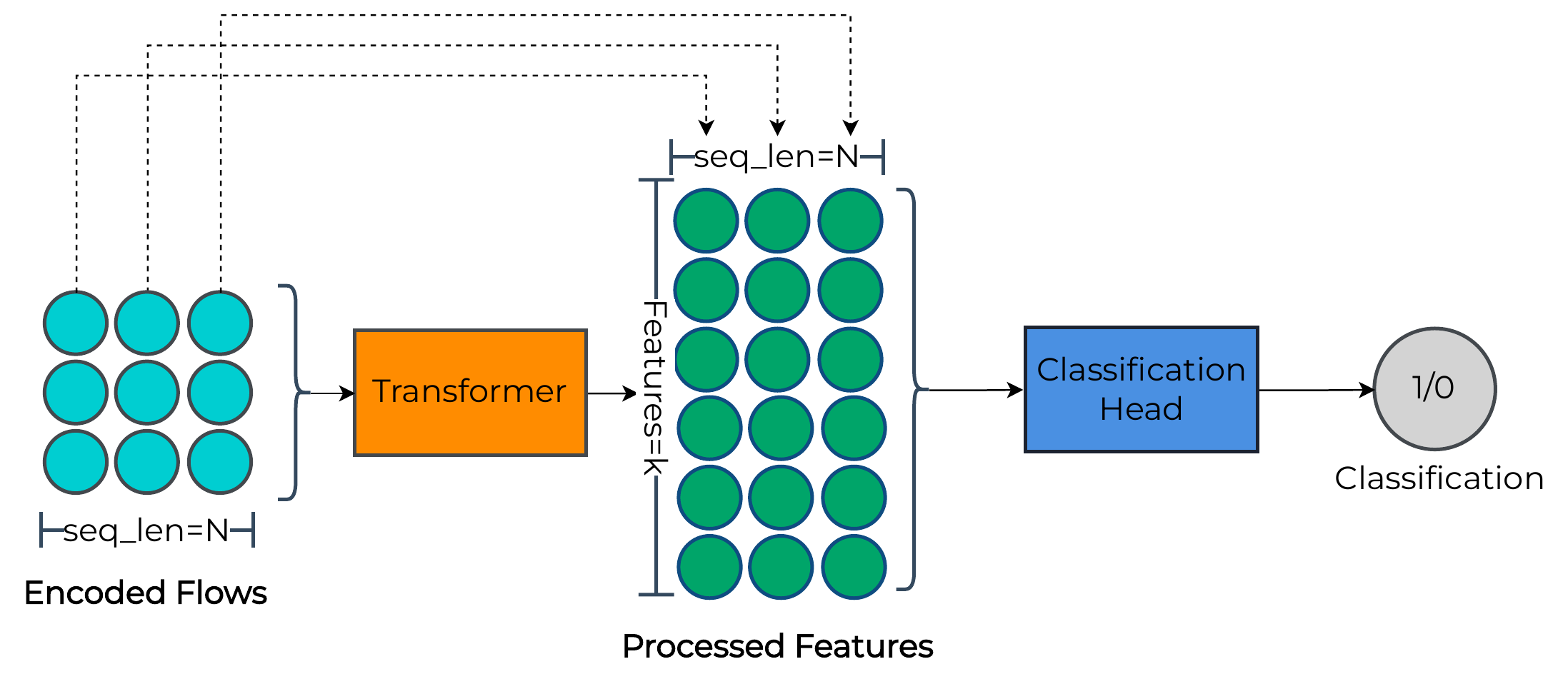}
    \caption{The classification head of a transformer model converts between the sequential output of the transformer, into a classification result.}
    \label{fig:Architecture_OutputHeadGeneric}
\end{figure}

We show the location and function of a classification head in Figure~\ref{fig:Architecture_OutputHeadGeneric}. Because transformers are sequence-to-sequence models, we must transform the sequential output into a classification result for an NIDS. Typically, this involves performing some operation on the output sequence, to prepare it to be passed into a dense Multilayer Perceptron (MLP), which then performs the final classification.
While it is possible to feed all the outputs from the transformer directly into the MLP, this causes the parameter count to rise exponentially as the sequence length increases, and therefore this is suboptimal in many applications. We call this approach 'flattening', as the feature vector is flattened before being passed to the MLP, and consider it in our evaluation. However, we also consider other, more efficient approaches, which will be discussed in this section.

The most common approach in the NLP domain is to average over the features in the sequence, called Global Average Pooling. For each feature in the resulting feature vector, we simply take the average of this value over the entire sequence. 
This approach works well for tasks such as sentiment classification, where the average of features over an entire sentence or paragraph can be combined. However, in the case of flows, if only the last flow in a sequence is being classified, averaging over the context for previous flows is not necessarily a sensible decision.

Instead of simply taking a global average across every flow in the sequence, we could use a densely connected neural network layer, for each feature in the contextual representation. This is a time distributed approach, and allows the model to apply different weights to the different flows in the sequence, for example, more heavily utilising the representation of the last flow. This is referred to as Featurewise Embedding or Featurewise Projection in this work.
This, however, still includes information from previous flows that is already considered by the transformer itself. Rather than doing this, we can simply take the last output from the transformer and use this as input to the classification head. This corresponds to the feature vector from the last flow. We refer to this as Last Token in this work.
Since typically, the classification task of an NIDS is only concerned with the class (e.g. benign vs malicious) of the last flow, using only the last contextual embedding vector should be sufficient.

\begin{figure}[!t]
    \centering
    \includegraphics[width=1\linewidth]{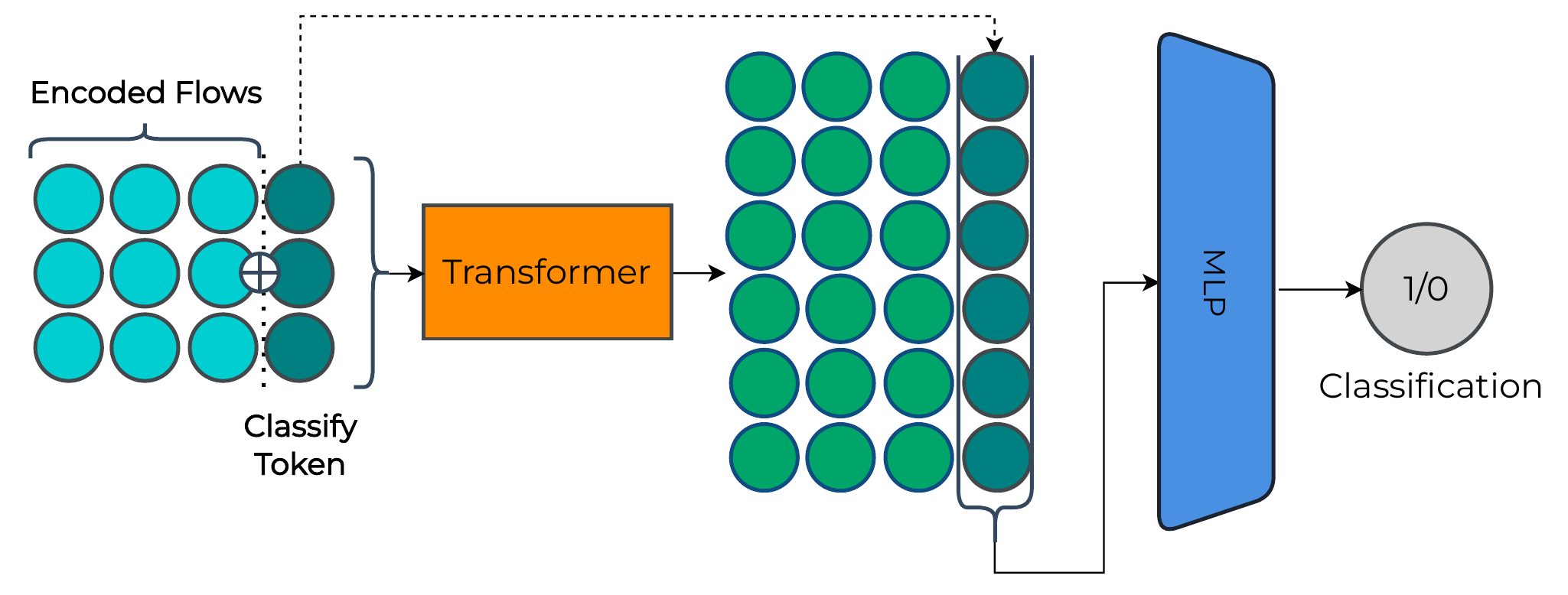}
    \caption{
    Here the classification head takes the output corresponding to the special 'classification' token, and feeds it directly into a dense network for classification.}
    \label{fig:Architecture_OutputHead_CLS}
\end{figure}

\begin{figure*}[!t]
    \centering
    \includegraphics[width=0.9\linewidth]{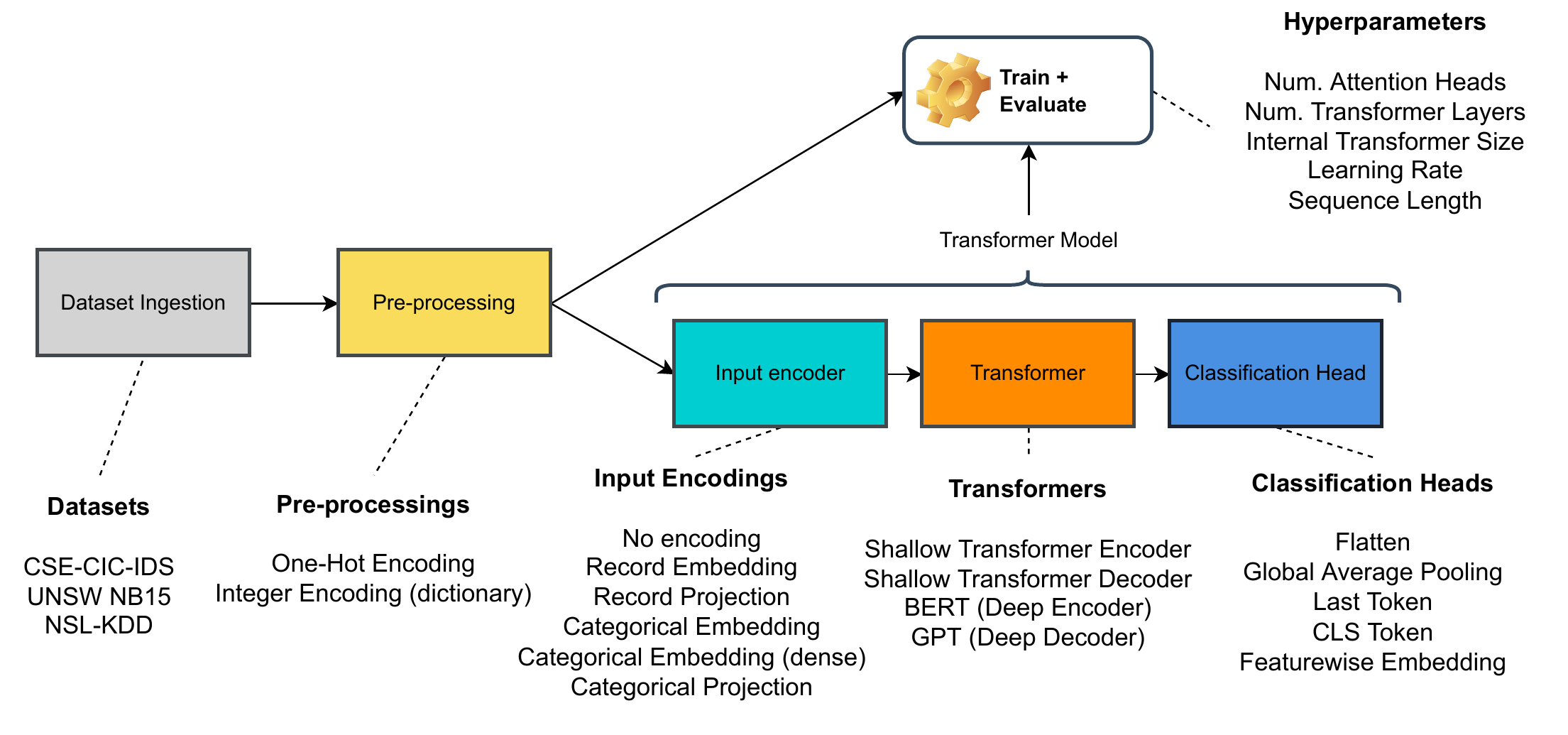}
    \caption{The FlowTransformer implementation allows the three primary aspects of a transformer to be interchanged, as well as different datasets and pre-processing approaches. Once a pipeline is specified, the evaluation is supported automatically, or a TensorFlow model can be exported for standalone training.}
    \label{fig:OverallLayoutFlowTransformerImpl}
\end{figure*}

Finally, models such as BERT, which are trained to perform multiple tasks such as masked language modelling and classification, introduce a special token that instructs the model to perform a particular task. 
We also test the efficacy of this in the NIDS domain, by evaluating the impact of appending a `classification' token to the end of a sequence of flows, and using the corresponding output of the transformer to perform classification. We give an example of this in Figure~\ref{fig:Architecture_OutputHead_CLS}. The advantages of this approach are that it can be used for a transformer system that is trained on both binary and multi-class classification, or for other tasks that do not fit into the typical classification scenario.

\section{Implementation}

The FlowTransformer framework implementation shown in Figure~\ref{fig:OverallLayoutFlowTransformerImpl} comprises several interchangeable blocks that form a processing pipeline. These blocks facilitate swift customisation, training, and evaluation of transformer models, and other sequential models, against flow-based NIDS datasets. In the sections below, we discuss each of these blocks.
We provide the source code on GitHub for public use and more detailed documentation, along with reference implementations for all the transformers used in this work \cite{ftgithub}.

\subsubsection{Dataset Ingestion}

FlowTransformer begins with a dataset ingestion component, which can accommodate any tabular dataset, making it compatible with various flow data formats. FlowTransformer automatically handles out-of-range or missing values. To apply FlowTransformer to a specific dataset, a dataset specification must be provided. This lists the categorical and numerical features of the dataset, for use by the pre-processing and input encoding stages. 
The dataset specification also identifies the class column to be trained on and the benign traffic label, both of which are used in the evaluation component to identify malicious and legitimate flows. Once the data specification is provided, FlowTransformer can ingest the dataset through the later stages without requiring coding.

\subsubsection{Pre-processing}

Upon receiving a dataset, FlowTransformer loads and divides it into training and evaluation splits for pre-processing. Users can select from three splitting configuration options for added flexibility. In this paper, we split the dataset in a 90\% to 10\% ratio. The pre-processing is first fit on the training data and then applied to the entire dataset, ensuring no information leakage from the evaluation dataset during pre-processing.
The pre-processing layer also takes into account the categorical format expected by the input encoder. Depending on the input encoder's requirements, the pre-processing layer can either one-hot encode or integer encode the categorical variables.
The pre-processing approach we employ in this work, is that proposed in \cite{lightbulb-nids}. It proved to be the most effective for neural network-based NIDS systems compared to other approaches tested. The pre-processing involves encoding the N most-frequent categorical features using one-hot or integer encoding (depending on the model's requirements) and min-max scaling numerical features after taking the logarithm.
Custom pre-processing can be specified by implementing fit and transform methods for numerical and categorical fields. The expected categorical format (one-hot or integer encoded) is provided as a parameter to the fit and transform methods for the categorical fields, allowing the pre-processing to be switched depending on the input encoding.

\subsubsection{Model Pipeline}

Once the pre-processing is complete, the input data's final size is determined, which is then used to construct the actual transformer model. FlowTransformer builds the transformer as a TensorFlow Keras model, which can be compiled and trained independently if required.

The transformer model comprises three interchangeable components, which we have discussed prior:

\begin{enumerate}
    \item The input encoder transforms the pre-processed data into an encoded format suitable for the transformer. These components can perform any transformation on the raw data. We provide one pre-processing implementation with the base framework, which we use for our evaluation.
    \item The transformer itself can be replaced with any transformer model, or any machine learning model that accepts a 3 dimensional input. We include implementations for basic transformer blocks, as well as transformer decoder blocks from GPT-3.0, and transformer encoder blocks from BERT. The number of flows in the sequence provided is fully configurable.
    \item The classification head receives the outputs from the transformer blocks and converts them into a classification. Although optimised for binary classification, multi-class classification heads are also supported within custom training loops when using FlowTransformer's built-in evaluation functions.\\ If needed, the classification head can also modify the tensor after input encoding, such as appending or removing tokens before passing them to the transformer, which is necessary for approaches like appending a classification token.
\end{enumerate}
    
\subsubsection{Evaluation Support}

FlowTransformer provides methods to evaluate a built model, offering several optimisations for faster dataset loads between model runs, enabling quicker hyperparameter searches. It also generates comprehensive outputs from each experiment, detailing parameters such as size, timing, performance results, and incremental results during training.
This functionality was utilised by our work for collecting results.

\section{Experimental Methodology}
\label{sec:ExpMethodology}

In this paper, we utilise the FlowTransformer framework to test a large number of applicable transformer components, to determine which are the most effective across a three benchmark NIDS datasets. We have previously discussed a number of input encodings, transformer blocks and output encodings, and each of these are tested using the FlowTransformer framework.

When collecting results, we performed a grid search. 
The dimensions explored in this paper are, input encodings, transformer block type, transformer depth, transformer feed forward size, number of attention heads, classification head and learning rate.

\subsection{Model Training and Grid Search}

When performing a grid search over a target space, we performed a minimum of 3 repeats of each experiment. We take the result from the best repeat, which can control for poor model initialisation during the experiments.
For the model training, we used early stopping and an epoch limit. The early stopping was set to a patience of 5 epochs, and the maximum number of epochs was limited to 20. We chose 20 epochs as we observed during our initial experimentation that the majority of models had converged to within 1\% of their final performance by the 20th epoch. We used the Adam \cite{kingma2017adam} optimiser for training with the parameters defined in the paper, with learning rate being specified in each experiment (as this was part of the grid search).

Results were collected with TensorFlow for Python 3.9 for Windows. TensorFlow was built and run with GPU support, on an NVIDIA GeForce RTX 2070 with Max-Q Design, and an Intel Core-i7-10750H CPU @ 2.6GHz with 6 cores.

\subsection{Datasets}

This paper considers three different widely used and highly cited NIDS datasets. Specifically, we use the flow format version of these datasets, which have become prevalent in the NIDS community. The conversion process of these datasets to flow format is detailed in \cite{Sarhan2021TowardsDatasets} which was proposed by the authors as a standardised format for flow-based NIDS.
\begin{enumerate}[align=right,itemindent=2em,labelsep=2pt,labelwidth=1em,leftmargin=0pt,nosep]
    \item \textbf{NSL-KDD}~\cite{ds-nslkdd}, an NIDS dataset for traditional networks that is one of the most widely used NIDS benchmark datasets. This was developed to replace the KDD cup datasets.
    \item \textbf{UNSW-NB15}~\cite{moustafa2015unsw}, an NIDS dataset for traditional networks featuring `a hybrid of real modern normal activities and synthetic contemporary attack behaviours' with `nine types of attacks, namely, Fuzzers, Analysis, Backdoors, DoS, Exploits, Generic, Reconnaissance, Shellcode and Worms. 
    \item \textbf{CSE-CIC-IDS2018} \cite{sharafaldin2018toward}, an NIDS dataset for traditional networks including `seven different attack scenarios: Brute-force, Heartbleed, Botnet, DoS, DDoS, Web attacks, and infiltration of the network from inside. 
\end{enumerate}

\subsection{Evaluation Metrics}

To assess the performance of different transformer models, standard metrics were utilised, such as F1 score, false alarm rate and detection rate. The metrics are computed using a combination of True Positives, True Negatives, False Positives, and False Negatives, denoted as $TP$, $TN$, $FP$, and $FN$ respectively. We use both F1 score as the primary metrics to compare approaches.


\subsection{Transformer Hyperparameters}

Besides comparing input encoding approaches and classification heads, we also explored various transformer configurations and corresponding hyperparameters. These are detailed in Table~\ref{tab:hyperparam}.

            \begin{table}[H]\small
            \renewcommand{\arraystretch}{1.3}
            \caption{Hyperparameter values used}
            \label{grid_search}
            \centering
            \scalebox{0.8}{ 
            \begin{tabular}{*3l }
                \toprule
                 
                 \textbf{Hyperparameter} & \textbf{Values}  \\
                \toprule
                
                Transformer Block & Encoder, Decoder\\
                Layers & 2, 4, 6, 8\\
                Feed Forward (FF) Dimensions & 128, 256, 512\\
                Attention Heads & 2, 4, 6, 8, 12\\
                Learning Rate & 0.01, 0.001, 0.0005, 0.0001, 0.00001\\
    
                \toprule

            \end{tabular}
            }
            \label{tab:hyperparam}
        \end{table}

\subsection{Inference \& Training Time}


For inference timing and training timing steps, we ensured that the GPU was cleared and the timings were begun from a warm TensorFlow state.
Training time was measured by timing each batch, which encompasses the backpropagation step. We then divided this by the batch size, and averaged over all batches used during training. This was done using TensorFlow's train\_on\_batch function.
We ensured that no extraneous processes were running during model training, and performed outlier testing to ensure that there was no significant drift of batch training times during training. 

To measure inference timing, we began by recording the inference time from the start to the end of a single batch. We repeated this process four times on the same batch of data to ensure the results were consistent and not being cached, and took the median time from those four measurements to obtain a stable range. We then repeated this procedure for 50 batches of data, ensuring that the batches were selected randomly, and calculated the mean time over all the batches.

Results are presented in terms of throughput, or the number of flows per second that can be ingested by the model.

\section{Results}

\begin{table*}[!t]
    \centering
    \includegraphics[width=0.85\linewidth]{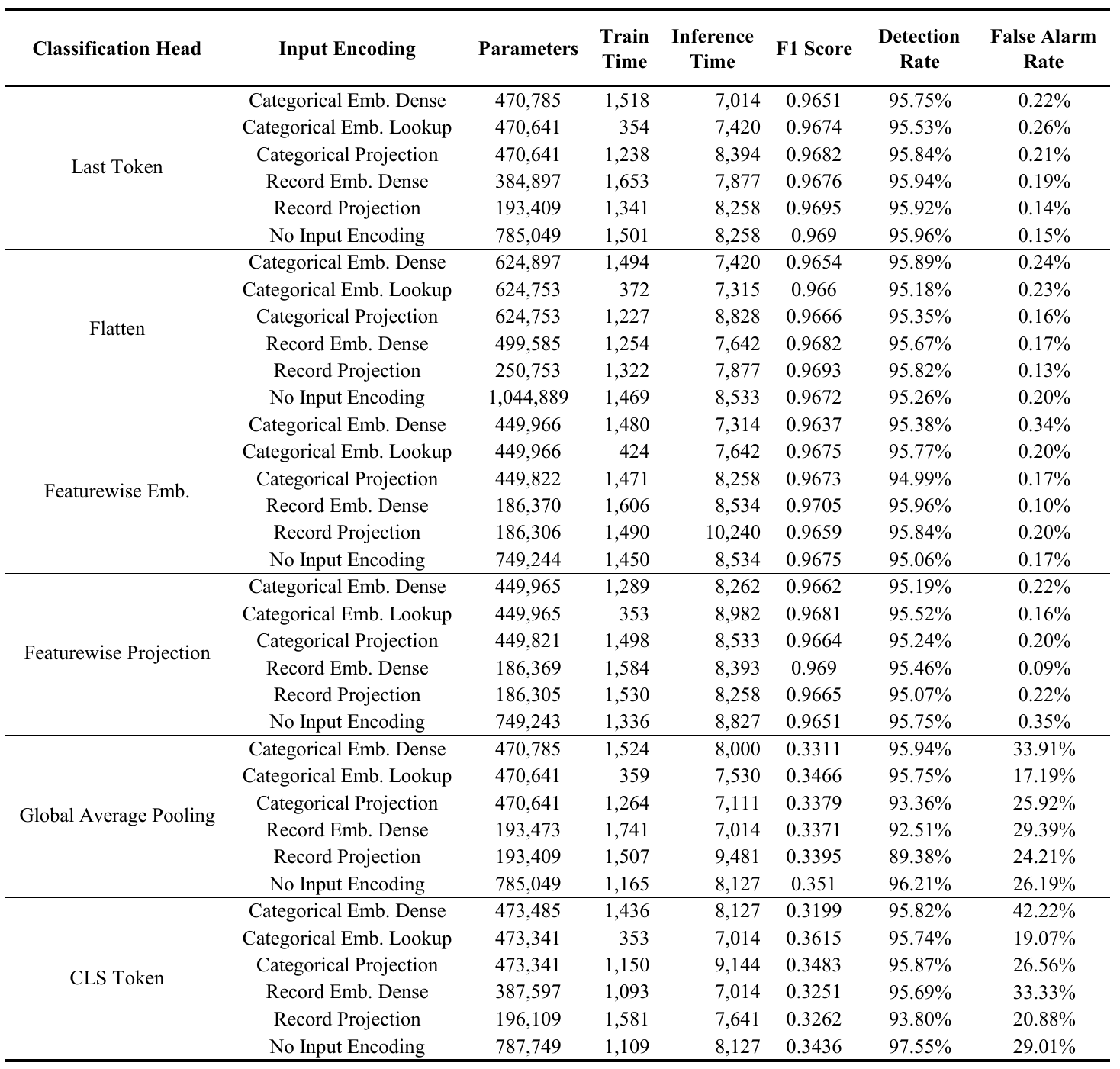}
    \caption{Results for a basic transformer model (2 layers, 2 heads, 128 ff. dim size), on the CSE-CIC-IDS2018 dataset. 
    }
    \label{tab:Table_InputAndHeadSummary_CSE}
\end{table*}

The results of this study are divided into four sections. The first section discusses the choice of input encodings, followed by a discussion of transformer blocks, classification heads, and finally, training and inference times. The study also evaluates five common transformer architectures, including BERT and GPT, by comparing the best choice of input encodings and classification heads for each of these models.


\subsection{Input Encodings}

\begin{figure}[!b]
    \centering
    \includegraphics[width=1\linewidth]{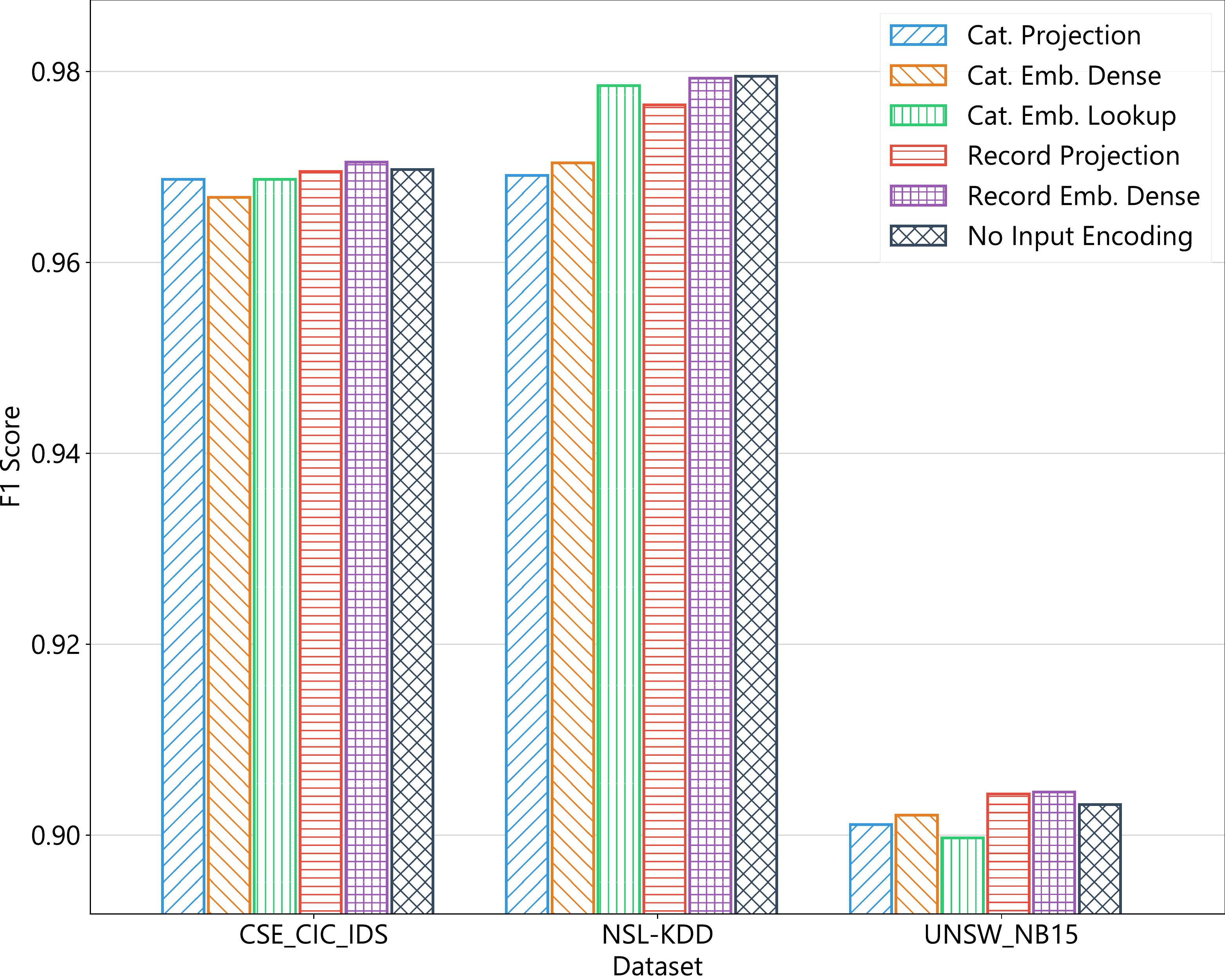}
    \caption{Results for various input encodings, for the basic transformer model (2 layers, 2 heads, 128 ff. dim size), on the CSE-CIC-IDS2018, NSL-KDD and UNSW-NB15 dataset.
    }
    \label{fig:groupbar_input_versus_f1}
\end{figure}


The impact of different classification heads and input encoding approaches on model size (number of parameters), detection performance (F1 Score, False Alarm Rate, True Positive Rate) and training and inference speed (throughput, flows per second) are presented in Table~\ref{tab:Table_InputAndHeadSummary_CSE}. The table shows the results for a basic transformer model (details below) trained on the CSE-CIC-IDS2018 dataset. The corresponding results for the NSL-KDD and UNSW-NB15 datasets are largely consistent with the results in Table~\ref{tab:Table_InputAndHeadSummary_CSE}, and are provided in the appendix, in Table~\ref{tab:Table_InputAndHeadSummary_NSL} and Table~\ref{tab:Table_InputAndHeadSummary_UNSW} respectively. The basic transformer model used in the experiment is a two-layer, two-attention head shallow transformer encoder, which is a sensible starting point for comparison, given its smaller size. Our evaluation found that these trends are consistent across all model depths, and that shallow transformers achieve competitive performance when compared to deeper models. Learning rates were chosen for each model through a grid search over the hyperparameter range listed in Section~\ref{sec:ExpMethodology}.

The comparison of different input encodings in terms of F1 score in Table~\ref{tab:Table_InputAndHeadSummary_CSE} indicates that no method significantly outperforms the others. This result is also presented in a graphical form in Figure~\ref{fig:groupbar_input_versus_f1} across the three datasets. The figure is an aggregate version of Tables \ref{tab:Table_InputAndHeadSummary_CSE}, \ref{tab:Table_InputAndHeadSummary_NSL}, and \ref{tab:Table_InputAndHeadSummary_UNSW}, and shows the performance of various input encodings. Each bar represents a single input encoding, while each group of bars represents a dataset. The y-axis represents the F1 score from the best choice of classification head for that particular input encoding.

Figure~\ref{fig:groupbar_input_versus_f1} clearly shows that the change resulting from various input encodings is minimal, and any choice of input encoding is still capable of achieving a reasonable model predictive performance.
However, when considering Table~\ref{tab:Table_InputAndHeadSummary_CSE}, it is clear that the choice of input encoding is one of the largest determining factors in regards to parameter count. Although categorical embedding is one of the most widely used approaches, record level embedding achieves the same performance with only half the model parameter count. For example, if we consider input encodings using the Last Token classification head in Table~\ref{tab:Table_InputAndHeadSummary_CSE}, categorical feature embedding has a parameter count of  $\approx$ 500K, and record level embedding has a parameter count of $\approx$ 200K.
This is particularly important for resource constrained systems.

Next, we look at the difference between projection and embedding.
We can see that both projection and embedding approaches performed equivalently, with projection sometimes outperforming embedding. Projection is a simpler linear transformation and embedding is non-linear. It could be argued that if a linear transformation of the input is sufficient to achieve the highest accuracy, then we could also use no input encoding and allow the model to learn through weights. However, a projection layer is a very effective form of dimensionality reduction, and can significantly reduce model size. It is also smaller than an embedding layer due to its lack of bias. 

The model parameter count can also be influenced by comparing the use of `categorical only' embedding versus `record level embedding'. Our experiments showed that both the overall record embedding (categorical and numerical fields combined) and embedding the categorical fields individually performed similarly in terms of accuracy. However, the former approach allows us to reduce the dimensionality across all features, rather than just the categorical fields, which can help reduce model size without sacrificing accuracy. This is particularly beneficial for flow formats that have a larger number of features or a substantial amount of numerical features that cannot be categorically embedded. 
In addition to a reduction in model size, the most significant advantage to these record level approaches comes when we look at training and inference time. 
In terms of embeddings, the most commonly found approach in the context of Transformers for NLP, is lookup-based embedding for the handling of categorical fields. This works by creating a mapping between the input levels and a vector which can be learned during training time, creating a dictionary lookup. This approach is typically highly scalable, and while we expect to see a slower training time for this approach, which we do observe, we can see in Table~\ref{tab:Table_InputAndHeadSummary_CSE} that record level approaches also outperformed lookup embedding in terms of the number of flows per second during inference.
This means in the NIDS domain, the record level approaches can result in smaller, faster training, and faster inference time models, without sacrificing performance, compared to other more commonly used approaches.

\subsection{Transformer Size}

We will now examine the transformer models themselves. We will analyse various aspects of the transformer construction, including the number of encoder blocks or layers, the internal feed-forward dimension, and the number of heads. To facilitate these comparisons, we will use transformer encoder blocks, as we did for the input encoding comparison.

\begin{figure}[b]
    \centering
    \includegraphics[width=1\linewidth]{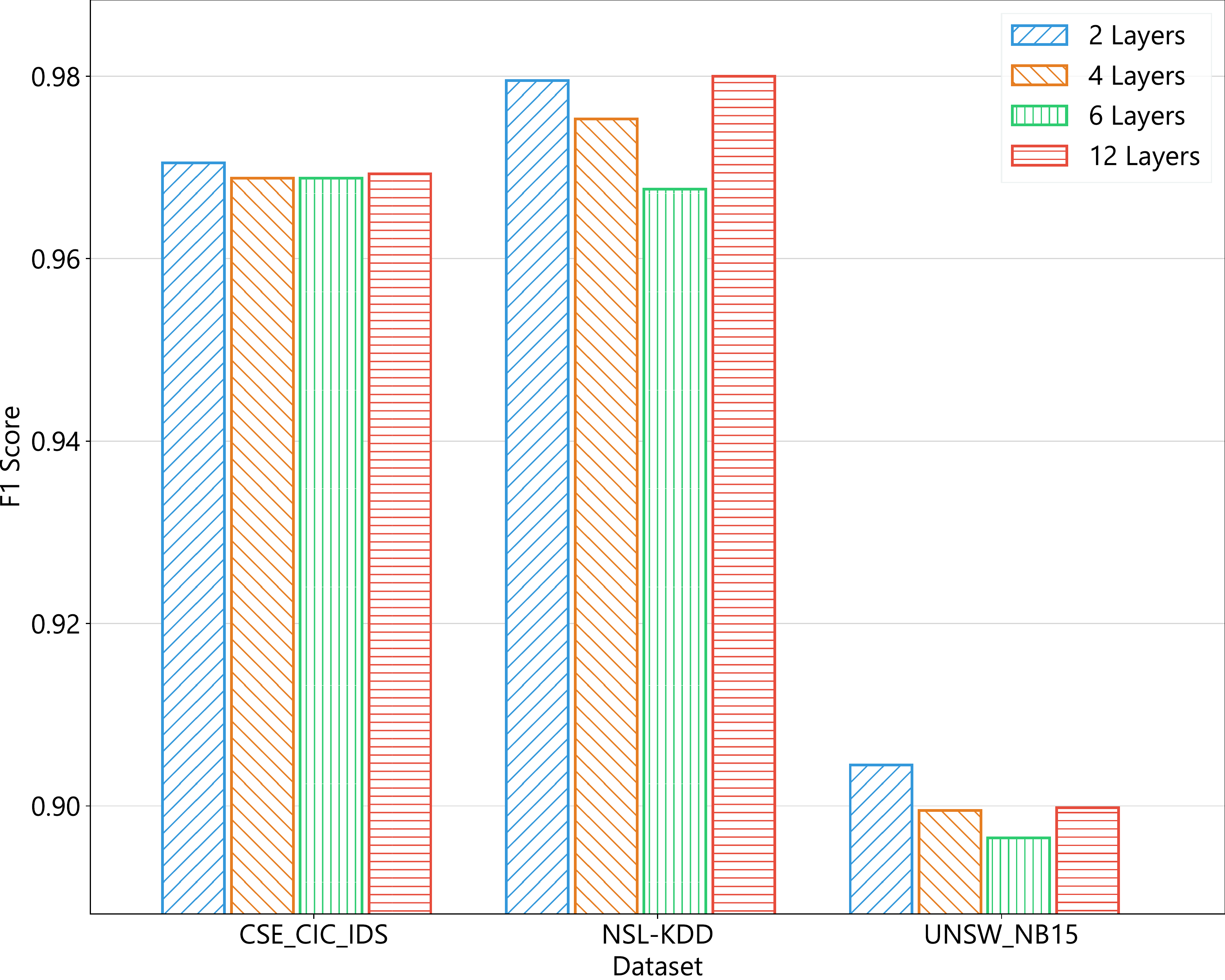}
    \caption{The F1 scores of four depths of transformer models, across the three benchmark datasets. 
    }
    \label{fig:groupbar_layers_versus_f1}
\end{figure}

Figure~\ref{fig:groupbar_layers_versus_f1} compares transformer decoder models with different numbers of layers, in terms of their F1 score across the three benchmark datasets. Each group of bars corresponds to one dataset, and each bar corresponds to a particular transformer with that number of layers. 
The results shown in the figure are for the best-performing models based on the input encoding, classification head, and other hyperparameters used. This ensures a fair comparison between the best models with a particular depth, internal size, and number of heads. 

\begin{figure}[!t]
    \centering
    \includegraphics[width=1\linewidth]{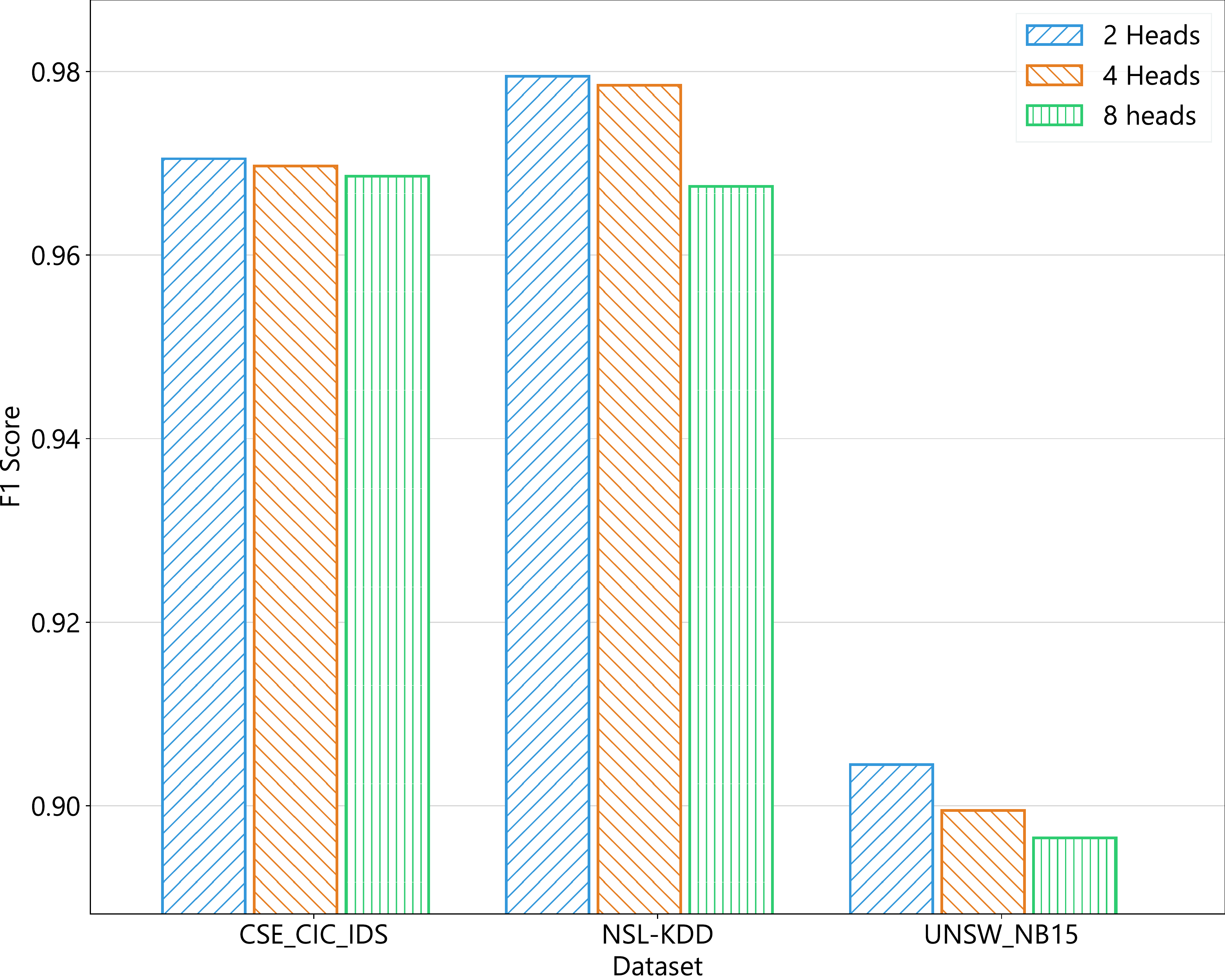}
    \caption{The F1 scores of transformer models with 2, 4 and 8 attention heads respectively, across the three benchmark datasets. 
    }
    \label{fig:groupbar_heads_versus_f1}
\end{figure}

\begin{figure}[!b]
    \centering
    \includegraphics[width=1\linewidth]{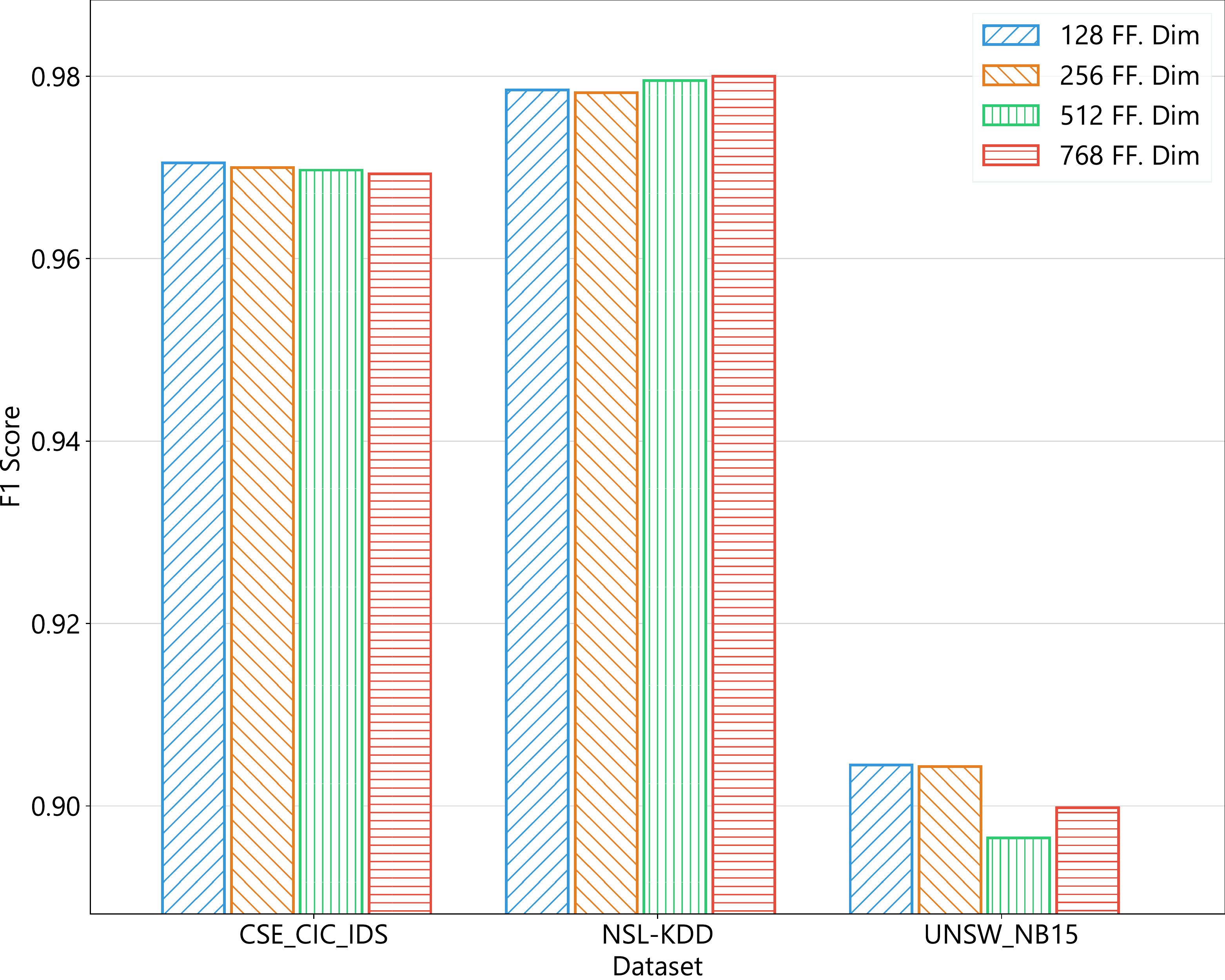}
    \caption{The F1 scores of transformer models with four different internal feed forward sizes, across the three benchmark datasets. 
    }
    \label{fig:groupbar_size_versus_f1}
\end{figure}

We can see in Figure~\ref{fig:groupbar_layers_versus_f1} that the model depth does not have a very large impact on the model performance. The shallowest 2-layer model outperformed the deeper models across all three datasets. This trend is mirrored in Figure~\ref{fig:groupbar_heads_versus_f1}, which compares the number of model heads, where again, the smallest 2-head model outperformed the models with more heads.
For the internal feed forward size, shown in Figure~\ref{fig:groupbar_size_versus_f1}, again the smallest internal size performs best for the CSE-CIC-IDS2018 and UNSW-NB15 dataset, but notably for the NSL-KDD dataset, the largest internal size performed the best. 

The number of attention heads, internal size and number of layers required for a task are all dependent on the complexity of the task, and the data available, so the variations between datasets are to be expected. However, across the 3 datasets we tested, shallow models showed comparable performance to deeper models. 
Where shallow models are applicable for a particular task, they can be preferable, as models with fewer parameters require less data to train, and are generally less prone to overfitting. 
Based on our observed results in the NIDS domain, shallow transformer models should not be ruled out in favor of larger models. Later we will include a specific comparison with two shallow transformer models, and two well-known large transformer models, GPT and BERT, to enable further comparison.

\subsection{Classification Heads}

When investigating the impact of classification heads, we refer back to Table~\ref{tab:Table_InputAndHeadSummary_CSE}, which shows the impact of different classification heads and input encodings on the F1 score, the number of model parameters, as well as  and training and testing time.
When comparing the F1 scores of different classification heads, it is apparent that classification heads had the largest overall effect on performance, and notably, the effect was consistent across all datasets.

\begin{figure}[!t]
    \centering
    \includegraphics[width=1\linewidth]{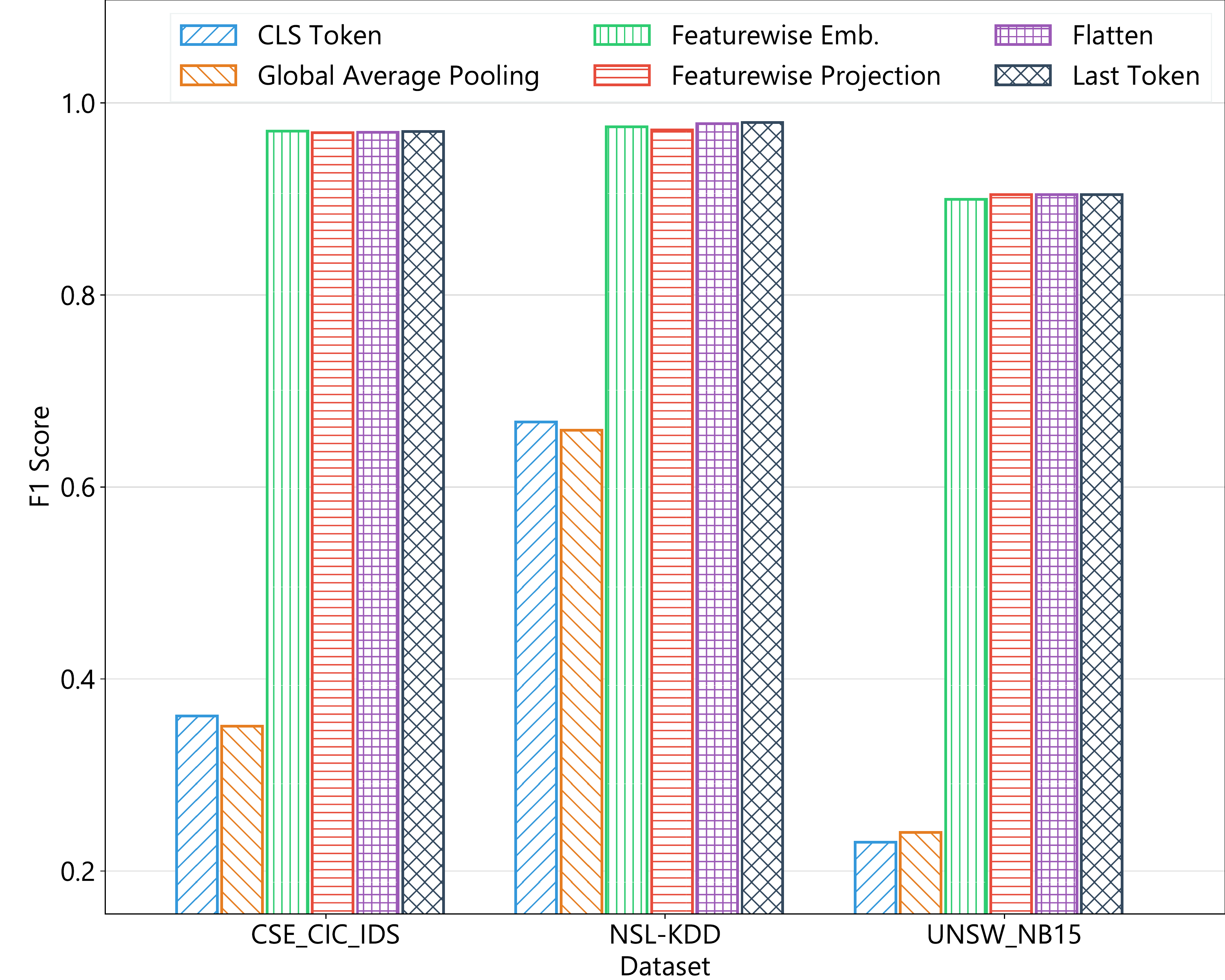}
    \caption{The F1 scores of transformer models with six different classification heads, across the three benchmark datasets. 
    }
    \label{fig:groupbar_classhead_versus_f1}
\end{figure}

To make this trend clear, we show the a graphical version of tables \ref{tab:Table_InputAndHeadSummary_CSE}, \ref{tab:Table_InputAndHeadSummary_NSL} and \ref{tab:Table_InputAndHeadSummary_UNSW} in Figure~\ref{fig:groupbar_classhead_versus_f1} for each of the three benchmark datasets. Each bar represents a choice of classification head, and the y-axis is the F1 score. We choose the F1 score from the model with the best choice of input encoding for this classification head.
This graph shows that the consistently best approach is using the last token classification head, followed by flatten, then the featurewise embedding approaches. However, each of these approaches has a relatively high performance, in contrast to CLS and Global Average Pooling, which performed very poorly. 
This is a significant result. A large number of related works,  particularly in the natural language processing domain, use Global Average Pooling when performing classification. However, this was one of the lowest performing approaches in the NIDS domain. This does make sense, since the NIDS context, only the last flow in the sequence is being classified, and the relations between this flow and previous flows are already captured by the transformer in the last token. By averaging, we include information from previous flows that may not be relevant to the classification task at hand.

Among the highest performing classification heads was the Flatten approach, which uses the full outputs from the transformer and passes them directly to the classification MLP. However, we can see in Table~\ref{tab:Table_InputAndHeadSummary_CSE} that this comes at the cost of a disproportionately large parameter count, which increases exponentially as the sequence length increases, unlike for the other approaches. 
Although for small sequences of flows this approach can be used, for longer sequences the model size can become too large. A larger number of model parameters requires more training data and increases the likelihood of overfitting.

Finally, featurewise embedding also performed relatively well. This approach allows a weighted combination from all of the features of flows in the sequence to be passed to the classification MLP, rather than just the features from the last flow. However, this is likely not advantageous over simply taking only the last flow's embedding, as the transformer has already captured dependencies between the last flow and previous flows. And given that this achieves a slightly lower F1 score than taking the last token, this means that in the NIDS domain, last token classification heads appear to be the most effective.

\begin{table*}[!t]
    \centering
    \includegraphics[width=1\linewidth]{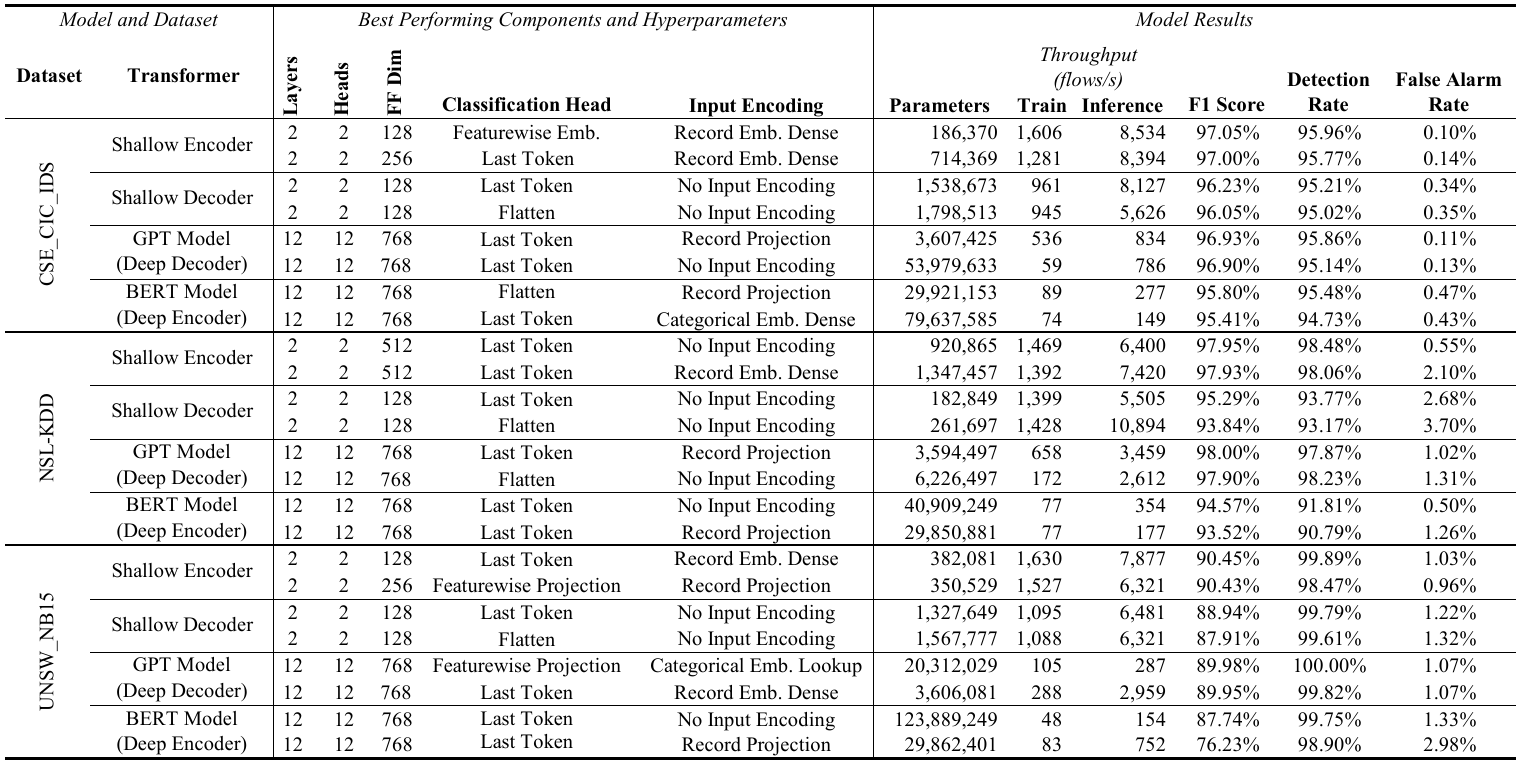}
    \caption{For each dataset and the four identified models, we display the results for the two highest performing combinations of parameters. These results are from the best choice of model hyperparameters, input encoding and classification head and learning rate for each dataset.}
    \label{tab:Table_AllModelsCompared}
\end{table*}

\subsection{Comparative Evaluation}

Finally, we provide a comparative evaluation of four transformer models. This evaluation covers the following models:

 \begin{itemize}
     \item Basic Dense Encoder - A two encoder block transformer
     \item Basic Dense Decoder - A two decoder block transformer
     \item GPT Model - A 12 decoder block transformer, with a 768 internal dimension and 12 attention heads 
     \item BERT Model - A 12 encoder block transformer, with a 768 internal dimension and 12 attention heads
 \end{itemize}

 The results are shown in Table~\ref{tab:Table_AllModelsCompared}. For each model, we include the two best combinations of input encoding and classification head, and the corresponding parameter count, training and inference times, F1 score, detection rate and false alarm rate.
 
We can see that both the shallow and deep models have a similar F1 score. However, the deep models have parameter counts orders of magnitudes larger than the shallow models. This model size difference is also reflected in the training and inference throughput, with the shallow models having a much higher throughput than the deep transformer models. With the CSE dataset, the best shallow encoder has $\approx$ 200K parameters with a throughput of $\approx$ 8.5K flows per second, whereas the deep encoder has $\approx$ 29M parameters and a throughput of just $\approx$ 300 flows per second. Because the accuracy of the deeper models is lower than those of the shallower models, it does not appear that GPT or BERT are optimal choices for network intrusion detection.
During our initial experimentation we also evaluated a subset of these larger models with long training time and no early stopping, and we found that regardless of training time, the deep models did not outperform the shallow transformer models. 

The fact that the decoder models which include GPT performed comparably to the encoder models, is also an interesting observation, as a decoder model is generally thought to be more suited to generative tasks. However, the attention mechanism in models such as GPT is auto-regressive, meaning it only considers previous flows when making a prediction. This is well suited to a flow based classification task, where we have a realtime stream of flows, and want to determine if the most recent flow is part of a suspicious sequence. On the other hand bi-directional attention, such as those used by BERT, would consider flows in both directions, which is irrelevant for the last flow in the sequence.

Overall, based on our results, we can observe that the highest performing model consists of the following components. These likely represent a good starting point for exploring transformer models for flow-based network intrusion detection. 

\begin{itemize}
    \item Input encoding: Record level projection
    \item Transformer block: Decoder
    \item Transformer internal size: 1-2x input size
    \item Transformer layers: 2
    \item Transformer heads: 2
    \item Classification head: Last token
\end{itemize}

The input encoding and classification head likely represent good starting choices for a variety of datasets and transformer models, with the specific transformer itself being a good candidate for modification. The learning rate must be determined for a particular choice of model, and this should be done across repeated iterations of model training, given the training volatility we observed. The FlowTransformer framework provides an efficient platform for this type of experimental exploration.

\section{Conclusion}




This paper presents FlowTransformer, a modular framework that allows the rapid implementation and evaluation of transformers for NIDS. The framework supports the combination of a variety of common transformer components, including input encodings, classification heads, and transformer constructions, and enables the formulation of training and inference pipelines for flow-based networking data, specifically in the context of NIDS.

We have demonstrated the utility and power of FlowTransformer via an extensive and systematic evaluation of a wide range of transformer model configurations, across 3 widely-used NIDS benchmark datasets.  

We also specifically explored the feasibility of GPT 2.0 and GPT 3.0-based technologies for NIDS, together with size and performance trade-offs across three common NIDS benchmark datasets.
The key findings from our experimental evaluation are:

\begin{itemize}
    \item Overall, the choice of classification head was the most critical factor in model performance, and the best choice of classification head was `Last Token'
    \item By using record level embedding or projection, the model size can be reduced by more than half, without reducing classification performance. Record level approaches were also the fastest in terms of both inference and training time.
    \item Shallow transformer models are sufficient for certain NIDS tasks, including those present in the benchmark datasets we tested.
    \item Transformer encoders and decoder blocks were both effective, but given the smaller size and higher applicability of encoder blocks, these are likely the best choice
\end{itemize}

Finally, we compared the performance of shallow transformer models, with that of the larger GPT and BERT architecture for the task of NIDS classification. 
We showed that although in the case of GPT, the larger model was able to reach an equivalent level of performance as the shallower models, because of the increased size and lower model throughput, these are likely not the optimal choices for most NIDS tasks.

We believe FlowTransformer can provide a basis for further explorations of the great potential of transformer-based approaches in the specific context of network intrusion detection, and possibly beyond that.

\twocolumn 
\printbibliography


\appendix
\begin{table*}[t]
    \centering
    \includegraphics[width=0.67\textwidth]{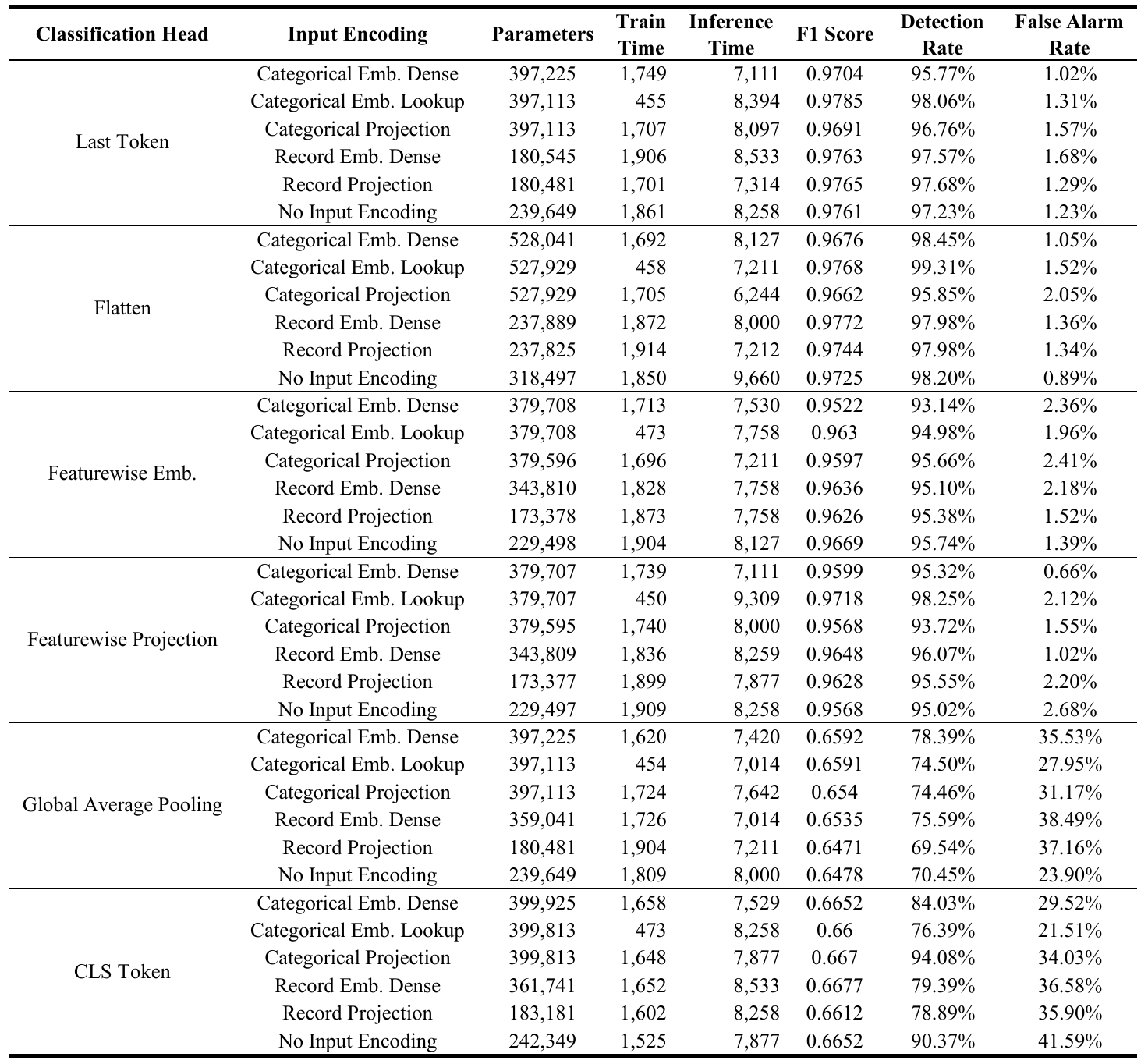}
    \caption{NSL-KDD Dataset, same format as Table~\ref{tab:Table_InputAndHeadSummary_CSE}}
    \label{tab:Table_InputAndHeadSummary_NSL}
    \includegraphics[width=0.67\textwidth]{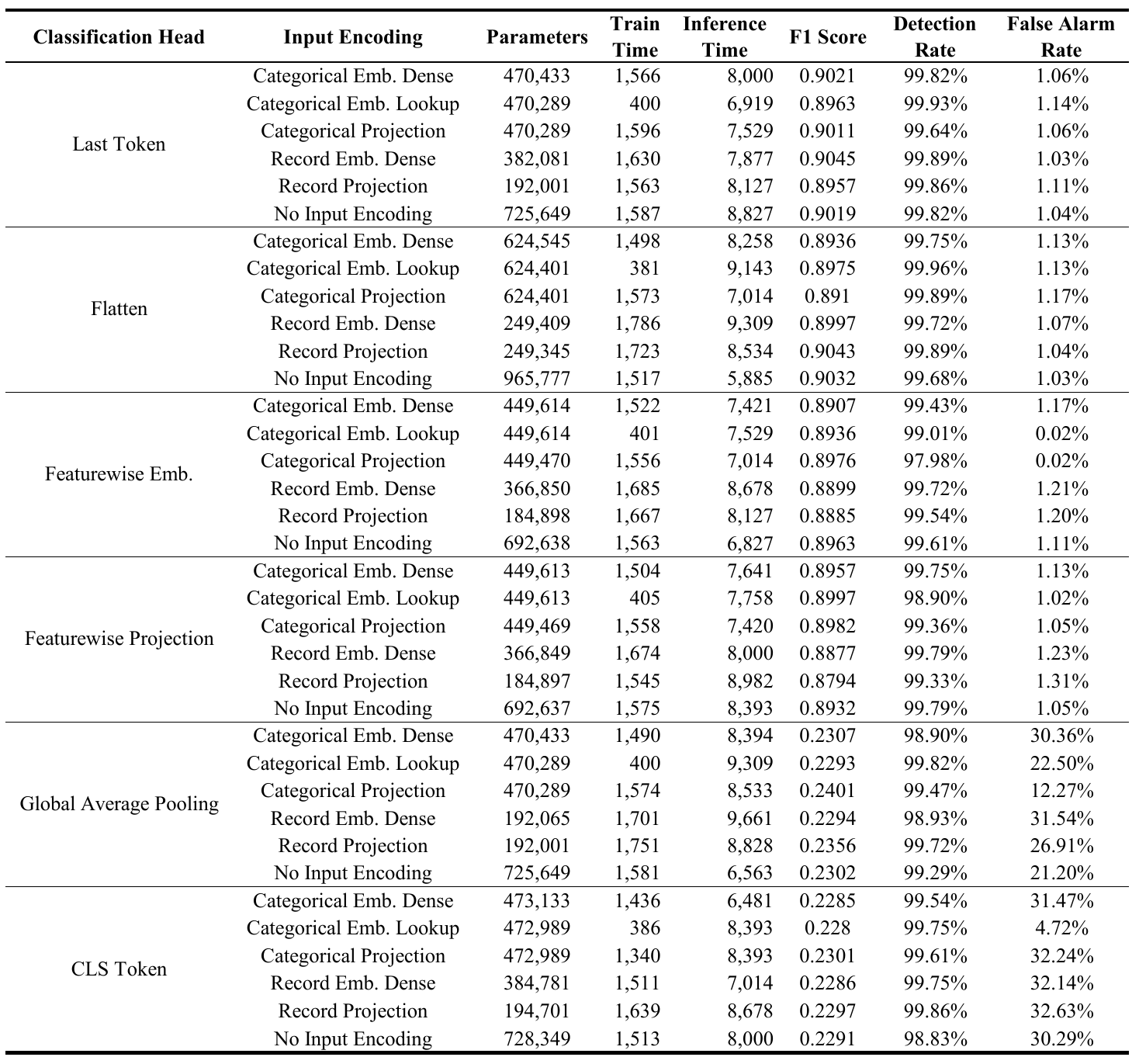}
    \caption{UNSW-NB15 Dataset, same format as Table~\ref{tab:Table_InputAndHeadSummary_CSE}}
    \label{tab:Table_InputAndHeadSummary_UNSW}
\end{table*}

\end{document}